\documentclass[twocolumn,twoside,slac_two]{revtex4}
\usepackage{graphicx}
\usepackage{fancyhdr}
\begin{document}

\title{Evolving Disks in Cold Dark Matter Cosmology}

%

\author{C. B. Brook$^1$, H. Martel$^1$}
\affiliation{1 Universit\'{e} Laval, Qu\'{e}bec, Qc, G1K-7P4 Canada}
\author{B. K. Gibson$^{2.3}$, D. Kawata$^2$}
\affiliation{2 Swinburne University, Vic., Australia}
\affiliation{3 School of Mathematical Sciences, Monash University, Australia}

\begin{abstract}
Despite having entered the era of ``precision cosmology,"  the formation of galaxies  within the favored  CDM cosmological paradigm remains problematic. By relating  our N-body/smooth particle hydro-dynamical simulation to an extensive range of Galactic and extragalactic observations, we  shed light on the formation and evolution of the Milky Way and other late type galaxies.  In light of recent observations of the stellar populations of extragalactic thick disks, we examine the proposal that the thick disk formed in a  high redshift period  characterized by gas rich merging. We show such a scenario to be consistent with color  observations. We then follow  the evolution of  structural parameters of the subsequently formed disk galaxy, from redshift $\sim$1 to the present. Consistent with observation, little evolution in the ratio of scale-height to scale-length is found in our simulated galaxy, despite its somewhat chaotic origins. The simulated galaxy  in this report forms part of a larger study of a suite of galaxies, with which  these issues are to be studied in  detail.
   
\end{abstract}

\maketitle

\thispagestyle{fancy}


\section{INTRODUCTION}
The age of precision cosmology is upon us, with concordance values of cosmological parameters placing us in an accelerating Universe, composed primarily of dark energy ($\sim$73\%), and dark matter ($\sim$23\%), with a mere smattering of baryonic matter ($\sim$4\%).  The evolution of structure on large scales is driven by gravity, and such structures as clusters and filaments of galaxies are exceedingly well explained by the $\Lambda$CDM paradigm. Yet on the scale galaxies, problems persist.  In some sense this is not surprising, as on smaller scales, baryonic processes become important;  gas dynamics, star formation, quasars, Active Galactic Nucleii, supernovae, shocks, and ram pressure are some of the processes that add complexity beyond gravity driven structure formation. Our work is aimed at unravelling the major processes which determine how galaxies form within the context of hierarchical structure build. 

The recently emerging  conviction that thick disks are prevalent in disk galaxies, and their seemingly ubiquitous old ages \cite{exampl-ref4}, means that  the formation of the thick disk, perhaps more than any other component, holds the key to unraveling the evolution of the Milky Way, and indeed all disk galaxies. In an earlier study \cite{exampl-ref1}, we
proposed that the thick disk was formed in an epoch of gas rich mergers, at high redshift. This hypothesis was based on comparing N-body/SPH simulations to a variety of Galactic and extragalactic observations, including stellar kinematics, ages and chemical properties.
Recently, the stellar populations in  four nearby edge-on disk galaxies were resolved \cite{exampl-ref7},  and it was confirmed that  thick disks are common features of spiral galaxies. Thick disk populations of all galaxies in this sample  are old and relatively metal rich, and  no significant color gradient with height above the plane ($\Delta$$V$$-$$I$/$\Delta$Z) is apparent. These observations will provide further constraints to our thick disk formation scenario, which we probe in  this study by examining how the ages and metallicties of our simulated thick disk stars vary with height above the plane.

Meanwhile, in a study of 34 edge-on disk galaxies in the Hubble Deep Field \cite{exampl-ref10}, selected for apparent diameter larger than 1.3'' and unperturbed morphology,  it was shown that distant and local disk galaxies are globally similar in their relative thickness or flatness, as characterised by the ratio of scale-height to scale-length ($h_{\rm{z}}$/$h_{\rm{l}}$). Disks at redshift $z\sim 1$ are smaller, in absolute value, than present day disk galaxies, and have a flatness ratio, $h_{\rm{z}}$/$h_{\rm{l}}$, only slightly larger, by a factor $\sim 1.5$. We examine the evolution of our simulated disk galaxy in the epoch which follows thick disk formation, in light of these observations, to determine whether a choatic thick disk formation epoch is consistent with a galaxy whose structural parameters do not evolve much since around redshift 1. 

The galaxy presented in this conference proceeding is the first of an ongoing study of a suite of simulated galaxies, which will probe deeper into the issues presented here, and   will be detailed in  \cite{exampl-ref2} \& \cite{exampl-ref3}.                   

\begin{figure*}[t]
\centering
\includegraphics[width=135mm]{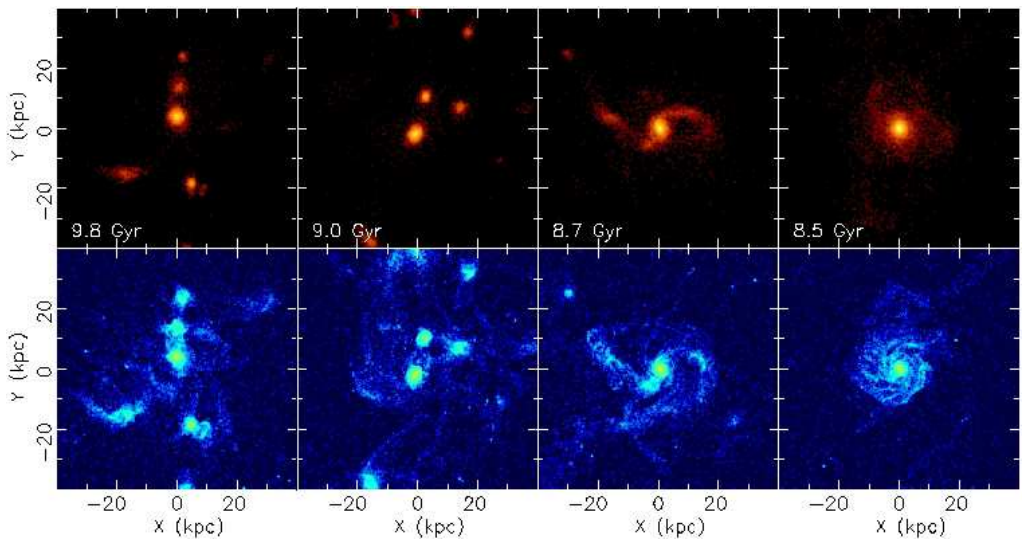}
\caption{Density plots of the evolution of the distribution of stars (upper panels) and gas (lower), shown face on, of the simulated galaxy between redshift $z=1.6$ and $z=1.1$.Axis are 100 kpc. This epoch is characterised by a series of merger events.} \label{f1}
\end{figure*}

\begin{figure}
\centering
\includegraphics[width=80mm]{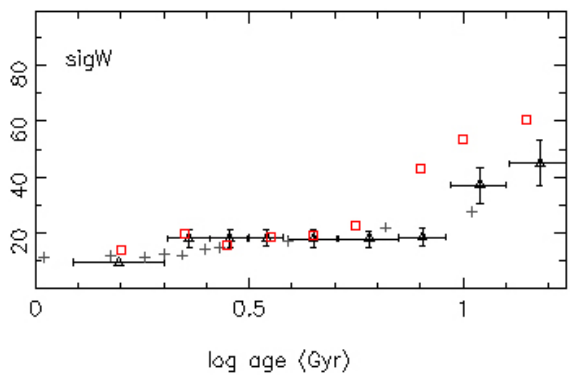}
\caption{Out of the plane direction (W) velocity dispersion versus log stellar age (Gyrs), plotted as red squares. Observations of solar neighbourhood stars are also shown, taken from  \cite{exampl-ref9} (triangles with error bars) and  \cite{exampl-ref8} (which selects for thin disk stars, ``+'' symbols). An abrupt increase in velocity dispersion at $\sim$10 Gyrs is apparent in the observations, and a similar abrupt increase is apparent in the simulated galaxy at $\sim$8.5 Gyrs.} \label{f2}
\end{figure}

\section{CODE AND MODEL}
We analyse  data from the simulated late type galaxy from a previous study \cite{exampl-ref1}. Our Galactic Chemodynamical code, ({\tt GCD+}), models
the effects of gravity, gas dynamics, radiative cooling, and star
formation in a self-consistent manner. Type~Ia and Type~II supernova feedback are included, and we trace the lifetimes of individual stars  when monitoring the 
chemical enrichment history of our simulated galaxies. 
Details of {\tt GCD+} can be found in  \cite{exampl-ref6}. 
We assume that 10$^{51}$ ergs is fed back as thermal energy from each SNe. An adiabatic phase is applied to gas heated by  SNe~II explosions.

We use the semi-cosmological version of {\tt GCD+},  with  an isolated sphere of dark matter and gas as initial condition.   Small scale density
fluctuations are superimposed on the sphere, parameterized by $\sigma_8$, which  seed local collapse and subsequent star
formation. Solid-body rotation is imparted to the initial sphere to incorporate the effects of longer wavelength fluctuations.  For the model described
here, the relevant parameters include the total mass ($5\times
10^{11}$~M$_\odot$), baryon fraction ($\Omega_{b}=0.1$), star formation efficiency, c$_*$=0.05, spin
parameter ($\lambda=0.0675$), and $\sigma_8=0.5$. We employed 38911 dark matter and 38911 gas/star particles.

\section{RESULTS}
Our final simulated galaxy has properties resembling the Milky Way. The galaxy has a dominant, exponential (scale-length$\sim$4.5 kpc), rotating disk comprising primarily young, metal rich stars. It also has a low mass, low metallicity, pressure supported stellar halo (for details see \cite{exampl-ref} , which examines the same simulated disk galaxy).

\subsection{Thick Disk Formation}

 The epoch of thick disk formation is depicted in Figure~\ref{f2}, which shows density plots of the evolution of the distribution of star particles (upper panels) and gas (lower), shown face on, of the simulated galaxy between redshift $z=1.6$ and $z=1.1$. Axis are 100 kpc. The velocity dispersion-age  plot (Figure~\ref{f1}) provides evidence that our simulated galaxy has a thick disk component, in close agreement with the velocity dispersion-age relation for solar neighbourhood stars \cite{exampl-ref9}, using  data of \cite{exampl-ref5}. The abrupt increase in velocity dispersion corresponds to end of the epoch depicted in Figure~\ref{f2}. This epoch,  is by far the most chaotic period in the galaxies evolution. At the beginning of this epoch, at least four proto-galaxies of significant mass exist. These building blocks are gas rich, with combined gas mass of $\sim$2.4$\times$10$^{10}$~M$_\odot$, compared with stellar mass of $\sim$7.0$\times$10$^{9}$~M$_\odot$. By $\sim$ 9 Gyrs ago, a single central galaxy has emerged.

\begin{figure}
\centering
\includegraphics[width=80mm]{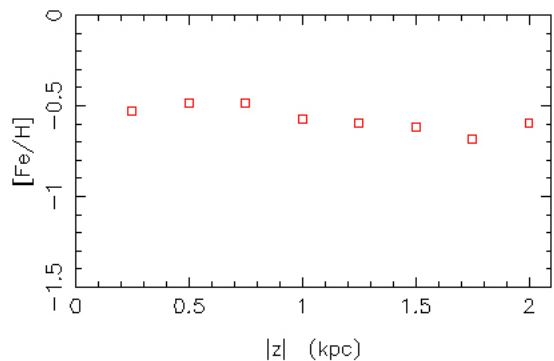}
\caption{the relation of the metallicity ([Fe/H]) of our simulated thick disk stars with height (kpc) above the plane. Very little gradient is apparent.  } \label{f3}
\end{figure} 

\begin{figure}
\centering
\includegraphics[width=80mm]{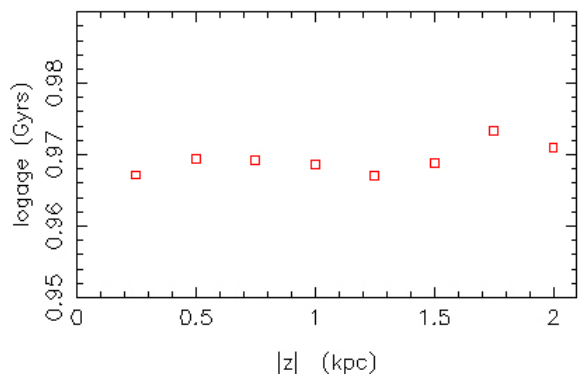}
\caption{log age versus height for the thick disk stars of our simulated galaxy. There is no gradient apparent. } \label{f4}
\end{figure} 

Our thick disk stars have metallicity [Fe/H]$\sim$-0.5 out to 2 kpc from the plane (Figure~\ref{f3}). The nature of thick disk formation ensures that our simulated thick disk stars are old, and that there is little variation in age with $|$Z$|$ (Figure~\ref{f4}).   

\begin{figure*}
\centering
\includegraphics[width=135mm]{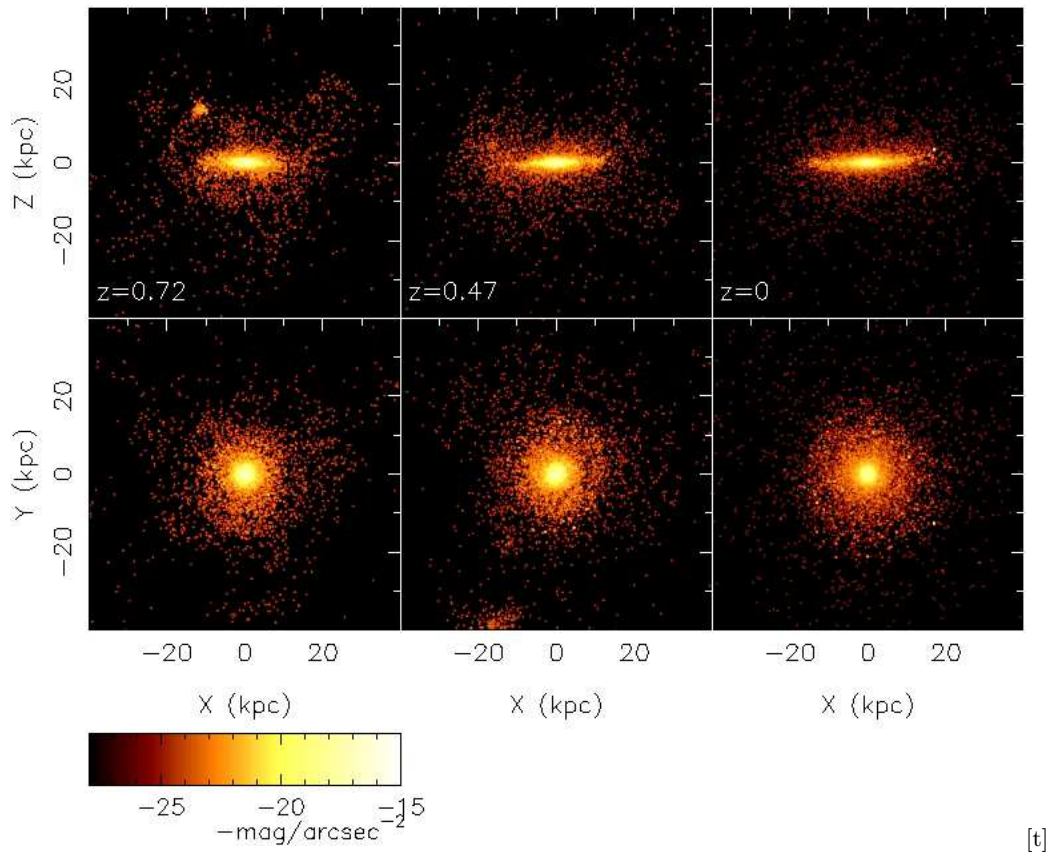}[t]
\caption{We zoom in to axis of 40 kpc, and show the $I$-band luminosity plots of our simulated galaxy, both edge-on (upper panels) and face on (lower panels). The flattened nature of the galaxy is apparent by the 1$^{\rm{st}}$ panel. The galaxy evolves relatively quiescently during the final 8 Gyrs.} \label{f5}
\end{figure*}

\begin{figure}
\centering
\includegraphics[width=75mm]{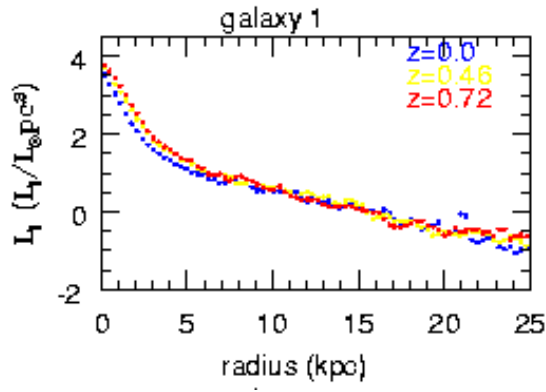}
\caption{$I$-band luminosity of the simulated galaxy versus radius, in the galactic plane,  taken at three redshifts. The earliest timestep ($z=0.72$, red) is taken after significant mergers have finished. An intermediate timestep ($z=0.46$, yellow) and present (blue) are also shown. The exponential nature of the disks between $5-15$ kpc allows scale-lengths to be derived. } \label{f7}
\end{figure} 

\begin{figure}
\centering
\includegraphics[width=75mm]{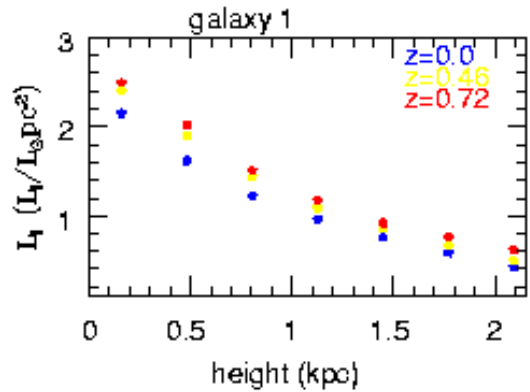}
\caption{$I$-band luminosity versus height above the plane. Symbols as for Figure~\ref{f7}. Exponentials are fitted between 0.5 and 1.5 kpc to derive scale-heights.} \label{f8}
\end{figure}

\subsection{Evolution of structural parameters}
The evolution of our simulated galaxy, in late epochs after the disk has formed, is shown in   Figure~\ref{f5}, where we plot  the $I$-band luminosity at three times,  $z=0.72$ (left panels), $z=0.46$ (middle), and $z=0$ (right), both edge-on (upper panels) and face-on (lower). 
The scale parameters of our disks can be measured by examination of radial and vertical luminosity profiles, shown in the $I$-band in Figure~\ref{f7} and Figure~\ref{f8}, respectively. It is evident that these parameters have not evolved much since the time that the disk galaxy has settled down. The luminosity surface density profiles   in Figure~\ref{f7} \& \ref{f8} are shown at redshifts $z=0.72$ in red, $z=0.46$ in yellow and $z=0$ in blue. We derive scale-lengths and scale-heights by fitting exponentials to these plots. The relative thickness to flatness can be parameterised by the ratio of scale-height to scale-length, $h_{\rm{z}}$/$h_{\rm{l}}$. This value is only larger by a factor of $\sim$1.2 in our simulated galaxy at $z=0.72$ compared with the present.

\subsection{Merger Histories}
\begin{figure}
\centering
\includegraphics[width=75mm]{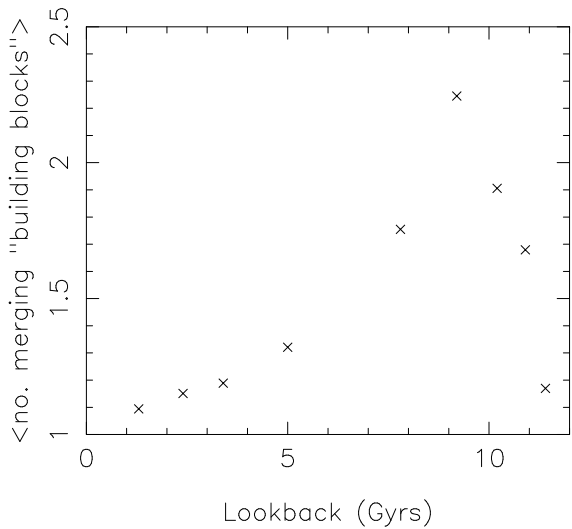}
\caption{The average number of building blocks versus lookback time (see text for details).}
\label{mergers}
\end{figure}

We make now a statistical study of merger histories in a  $\Lambda$CDM model with $\Omega_0=0.27$, $\lambda_0=0.73$, and $H_0=71\,\rm km\,s^{-1}Mpc^{-1}$,
using a $\rm P^3M$ algorithm  \cite{efstath}, \cite{hugo}. In our initial run, we simulated a comoving cubic volume of size
$L_{\rm box}=100\,\rm Mpc$, using $128^3$ particles. The mass per particle
was $1.801\times10^{10} {\rm M}_\odot$. We identified
a region of size $\rm 30\,Mpc\times35\,Mpc\times28\,Mpc$
in which several Milky-Way-like halos formed. We zoom-in by increasing
the resolution of the code by 2 in each directions in the region of interest
and rerunning the simulation. This simulation  has 
2.5~million particles, with mass per particle in the region of interest
of $2.251\times10^9 {\rm M}_\odot$. This is sufficient to resolve the relatively large building blocks which we are interested in. We have 58 halos with masses in the range $(0.5-1.1)\times10^{12} {\rm M}_\odot$. We plot, in Figure~ \ref{mergers},
the average number of ``building blocks'' versus lookback time for our 58 Milky Way sized dark matter halos. To qualify as a  ``building block'', the merging sub-halo must be $\ge10^{10} \rm{M}_\odot$, add at least $4\%$ to the mass of the halo, and be merged to the largest (central) halo by the next timestep. Thus, this plot summarises the history the build up of  Milky Way sized halos (see  \cite{exampl-ref2} for details). A peak merging period around 10 Gyrs ago, before a relatively quiescent final 7-8 Gyrs, is indicated by our plot, and provide support to the disk galaxy evolution scenario presented here.

\section{DISCUSSION}
The properties of disk galaxies have proven to be difficult to accommodate within the favoured Cold Dark Matter cosmological paradigm. Complex gas processes involving energy feedback, for example from quasars, supernovae and stellar winds, and multiple gas phases, have proven difficult to model. yet these processes appear to be crucial in the formation and evolution of disk galaxies within the context of hierarchical structure assembly. Here we report on a simulated disk galaxy, and match to recent observations of local and distant ($z\sim 1$) disk galaxies. In forming our simulated galaxy, feedback and gas physics proved crucial ingredients \cite{exampl-ref} in forming a realistic disk galaxy. We now have have a suite of four simulated disk galaxies, and the galaxy presented here will form an extensive study of disk galaxy formation and evolution (\cite{exampl-ref2} \& \cite{exampl-ref3} ) as we build up a statistical sample with which to interpret observations.

The simulated disk galaxy is shown to have a thick disk component. This is evidenced by the velocity dispersion versus age relation for solar neighbourhood stars (Figure~\ref{f1}). Our simulations indicate that the thick disk is created in an epoch of of multiple mergers of gas rich building blocks, during which a central galaxy is formed  \cite{exampl-ref1}, and it was shown that such a scenario is consistent with present observations of the Galactic and extragalactic thick disks. Recently,  observations of four edge-on galaxies explored properties of their stellar populations as a function of height  \cite{exampl-ref7}. All showed thick disks composed of red stellar populations, which are old and metal rich. $V$$-$$I$ colour gradients were seen to be zero or slighlty positive. These observations do not support thick disks which form through dissipative disk formation, accretion from shredded satllites, or slow heating of the thin disk. Heating through galaxy interactions more than 3 Gyrs ago are not ruled out. In Figures~\ref{f3} \& \ref{f4}, we show the metallicity and age of the thick disk stars in our simulated galaxy. We see that our thick disk formation scenario results in stellar populations which can match the observations. Although accretion from galactic ``building blocks'' are involved in our model, these accreted building blocks have high gas fraction. Astration during and after murders, as well as ongoing disspiative infall of gas, eliminates differences in metallicity of original and accreting material. Thus little metallicity gradient is seen in our simulated thick disk. The rapid star formation in  this epoch ensures that thick disk stars are almost exclusively old, and a lack of gradient with height.    

In a study of 34 edge-on disk galaxies in the Hubble Deep Field \cite{exampl-ref10},  galaxies out to redshift$\sim$1 were shown to have scale parameters similar to  local disk galaxies. These parameters are  characterised by the ratio of scale-height to scale-length ($h_{\rm{z}}$/$h_{\rm{l}}$). Disks at redshift $z\sim 1$ are smaller, in absolute value, than present day disk galaxies. However, their  flatness ratio, $h_{\rm{z}}$/$h_{\rm{l}}$, is only slightly larger, by a factor $\sim$1.5. We examine our simulated disk galaxy from a time after the thick disk formation epoch, when a late type galaxy has clearly formed. We derive a value of  $h_{\rm{z}}$/$h_{\rm{l}}$ which is larger by a factor $\sim 1.2$ in our simulated galaxy at $z=0.72$ compared with the $z=0$. Thus, even though a chaotic period of merging which led to the emrergence of the thick disk was involved in the birth of our simulated disk galaxy, once a disk emerges, the global structure does not evolve to a large extent. We found the disk does get larger and slighlty flatter.

\bigskip 
\begin{acknowledgments}
We thank Jeremy Mould for providing a preprint of his study on extragalactic thick disks.
CB and HM are supported by the Canada Research Chair program and
NSERC. BKG and DK acknowledge the financial 
   support of the Australian Research Council through its Discovery 
   Project program.

\end{acknowledgments}

\bigskip 

\end{document}